\documentclass[doublecol]{epl2}

\usepackage{graphicx}
\usepackage{dcolumn}
\usepackage{bm}
\usepackage{epsf}
\usepackage{amsmath}
\usepackage{amssymb}
\usepackage{color}

\newcommand{\bra}[1]{\left\langle #1\right|}
\newcommand{\ket}[1]{\left|#1\right\rangle}

\newcommand{\la}{\left\langle}
\newcommand{\ra}{\right\rangle}
\newcommand{\pd}{\partial}
\newcommand{\de}[1]{\delta\left(#1\right)}
\newcommand{\td}{\mathrm{d}}

\newcommand{\e}[1]{\exp{\left(#1\right)}}
\newcommand{\lo}[1]{\ln{\left(#1\right)}}

\newcommand{\com}[2]{\left[#1,\,#2\right]}

\newcommand{\bla}{bla\\bla\\bla\\bla\\bla}
\newcommand{\PR}{Phys. Rev.}
\newcommand{\PRA}{Phys. Rev. A}

\newcommand{\PRD}{Phys. Rev. D}
\newcommand{\PRE}{Phys. Rev. E}
\newcommand{\PRL}{Phys. Rev. Lett.}
\newcommand{\EPL}{EPL (Europhys. Lett.)}
\newcommand{\RMP}{Rev. Mod. Phys.}

\newcommand{\mc}[1]{\mathcal{#1}}

\newcommand{\mrm}[1]{\mathrm{#1}}

\title{Quantum entropy production in phase space}

\author{Sebastian Deffner}
\institute{Department of Chemistry and Biochemistry and Institute for Physical Science and Technology, University of Maryland, 
College Park, Maryland 20742, USA}
\date{\today}


\abstract{A fluctuation theorem for the nonequilibrium entropy production in quantum phase space is derived, which enables the consistent thermodynamic description of arbitrary quantum systems, open and closed. The new treatment naturally generalizes classical results to the quantum domain. As an illustration the harmonic oscillator dragged through a thermal bath is solved numerically. Finally, the significance of the new approach is discussed in detail, and the phase space treatment is opposed to the two time energy measurement approach.}

\begin{document}

\maketitle

Recent experimental progress put nanotechnological devices in the reach of realistic applications. Systems at the nanoscale, however, are subjected to both, thermal fluctuations and quantum effects, and they operate generically far from thermal equilibrium \cite{kinoshita_2006}. Especially for thermodynamic applications, as, for instance, the manipulation of ultracold gases in optical lattices \cite{bloch_2005} or the operation of quantum heat engines \cite{abah_2012}, microscopic properties of thermodynamic quantities as work, heat, or entropy production have to be understood.

Rather recently new insight has been gained by the discovery of the so called fluctuation theorems. Amongst the first, for small, but classical systems undergoing isothermal processes Jarzynski showed \cite{jar97}, that $\la \e{-\beta W}\ra=\e{-\beta \Delta F}$, where $W$ is the work, $\beta$ the inverse temperature of a surrounding heat bath, and $\Delta F$ the free energy difference. The angular brackets $\la\dots\ra$ denote an average over an ensemble of realizations of the same process, whose outcome is subject to thermal noise. The latter equality is generally valid for open and closed systems, slow (equilibrium) and fast (nonequilibrium) processes. The Jarzynski equality can be used to define the nonequilibrium entropy production, $\Sigma_\mrm{cl}$, as from the convexity of the exponential we have $\Sigma_\mrm{cl}\equiv\beta\left(\la W\ra-\Delta F)\right)\geq 0$ \cite{jar97}. Jarzynski's findings have been generalized to systems evolving under non-conservative forces \cite{hat01}, and it has been shown that fluctuation theorems can be understood as a consequence of the normalization of the probability distribution describing the state of the system under consideration \cite{sei05,sha10,seifert_2012}. All classical fluctuation theorems have in common that the entropy production $\Sigma$ is defined along a trajectory in phase space, for which generalizations to the quantum domain are not obvious. 

Therefore, to generalize the Jarzynski equality to quantum systems a different approach was pursued. For isolated systems  the work can be identified with the change in internal energy. By performing projective energy measurements in the beginning and in the end of a unitary quantum process, a natural definition of quantum work arises \cite{kur00,tas00}. The such defined work, however, is not an observable since the Hamiltonian does not have to commute with itself at different times \cite{tal07}. The two time measurement approach has been extended to describe processes with, for instance, heat exchange between two systems \cite{woj04}, or quantum systems weakly coupled to their thermal environment \cite{def11}, see also a recent review on the topic \cite{cam11}. 

The fluctuation theorems derived within the two time energy measurement approach, however, are not as general as their classical equivalents. Physically, energy measurements are only reasonable for isolated systems. For open systems the energy of the thermal reservoir has to be determined, so that distinguishing work and heat becomes feasible. If only partial information is available or the system of interest is not in a Gibbs state projective measurements inevitably lead to a back action of the measurement on the system \cite{kafri_2012}, and therefore the fluctuation theorem has to be modified. Thus, generalizations of the Hatano-Sasa relation \cite{hat01} or the Seifert fluctuation theorem \cite{sei05} to open quantum systems are still an open problem.

The present discussion starts by briefly opposing the classical Jarzynski equality and generalizations to the quantum domain, before we derive our main result. We prove a quantum fluctuation theorem valid for open and closed quantum systems undergoing  processes arbitrarily far from equilibrium and with arbitrary stationary states, equilibrium and nonequilibrium. To this end, we use the Wigner representation of the density operator $\rho$, for which a natural generalization of the classical expression for the entropy production becomes evident. To illustrate our findings we numerically solve the time-dependent harmonic oscillator coupled to a thermal environment, for which the exact master equation is known \cite{hu92,fleming_2011}. Finally, the paper is concluded by discussing the universality and significance of the new approach.

\section{Classical Jarzynski equality}

Consider a classical system, whose energy is described by the Hamiltonian $H(\Gamma,\alpha)$. Here, $\Gamma$ is a point in phase space, and $\alpha$ is a parameter that can be controlled externally. If the Hamiltonian is varied according to a protocol $\alpha_t$ the work performed during the time interval $t\in[0,\tau]$ reads \cite{jar97} 
\begin{equation}
\label{q01}
W_\mrm{cl}\left[\Gamma_\tau;\alpha_\tau\right]=\int_0^\tau\td t\,\dot{\alpha}_t\,\pd_\alpha H(\Gamma_t,\alpha_t)\,.
\end{equation}
Note that generically the so defined work depends on the whole trajectory in phase space, $\Gamma_t=(x_t,p_t)$. Its statistics are given by the probability density 
\begin{equation}
\label{q02}
\mc{P}_\mrm{cl}(W)\equiv\la \de{W-W_\mrm{cl}\left[\Gamma_\tau;\alpha_\tau\right]}\ra\,,
\end{equation}
where the angular brackets denote an average over all possible realizations. If the system is initially in an equilibrium state, $p(\Gamma_0,\alpha_0)=\e{-\beta H(\Gamma_0,\alpha_0)}/Z(\alpha_0)$, with respect to inverse temperature $\beta$, it follows \cite{jar97,jar97a,speck_2007} that the irreversible entropy production obeys an integral fluctuation theorem, $\la \e{-\Sigma_\mrm{cl}}\ra =1$, where for isothermal processes $\Sigma_\mrm{cl}=\beta\la W_\mrm{ir}\ra=\beta \left(W-\Delta F\right)$ and $F(\alpha)=-\beta\ln{Z(\alpha)}$. The latter equality is a theorem which remains true under fairly general conditions. In particular, it is applicable to isolated (Hamiltonian) as well as open (stochastic) system dynamics. Moreover, the processes under consideration can operate arbitrarily far from equilibrium. However, there are two conditions which have to be met: (i) the system is  initially prepared in a Boltzmann-Gibbs equilibrium state corresponding to the temperature of the environment, and (ii) for all values of $\alpha$ a stationary solution of the dynamics is given by the Boltzmann-Gibbs distribution, $p_\mrm{stat}(\Gamma,\alpha)=\e{-\beta H(\Gamma,\alpha)}/Z(\alpha)$. 

For systems evolving under non-conservative forces the The irreversible work, $\beta\la W_\mrm{ir}\ra$ is to be replaced by the nonequilibrium entropy production \cite{hat01,deffner_2011},
\begin{equation}
\label{q04}
\Sigma_\mrm{cl}\left[\Gamma_\tau;\alpha_\tau\right]=-\int_0^\tau\td t\,\dot{\alpha}_t\,\pd_\alpha \varphi(\Gamma_t,\alpha_t)\,, 
\end{equation} 
where $\varphi(\Gamma,\alpha)$ is the potential of mean force defined by $p_\mrm{stat}(\Gamma,\alpha)=\e{-\varphi(\Gamma,\alpha)}$ describing a \textit{nonequilibrium stationary state}. 

\section{Quantum Jarzynski equality}

In generalizing the fluctuation theorems to quantum systems one encounters the difficulty that the classical notion of a trajectory is not directly applicable. For thermodynamic purposes a way around this obstacle was found for \textit{isolated} dynamics by identifying work with an energy difference. To this end, imagine the following protocol: first a projective energy measurement is performed on the initial state $\rho_0=\e{-\beta H(\alpha_0)}/Z(\alpha_0)$; then the system undergoes unitary dynamics, $\rho_t=U_t\rho_0U^\dagger_t$, generated by the externally controlled Hamiltonian $H(\alpha_t)$;  finally a second projective energy measurement is made on the final state $\rho_\tau$. The work for one such 'realization' is given by \cite{kur00,tas00}
\begin{equation}
\label{q05}
W_\mrm{qm}[\ket{m(\alpha_\tau)};\ket{n(\alpha_0)}]=E_m(\alpha_\tau)-E_n(\alpha_0)\,,
\end{equation}
where $\ket{n(\alpha_0)}$ is the initial energy eigenbasis with eigenenergies $E_n(\alpha_0)$ and $\ket{m(\alpha_\tau)}$ the final basis with corresponding eigenvalues $E_m(\alpha_\tau)$. Accordingly, the work distribution takes the form
\begin{equation}
\label{q06}
\mc{P}_\mrm{qm}(W)=\sum_{m,n}\de{W-W_\mrm{qm}[\ket{m(\alpha_\tau)};\ket{n(\alpha_0)}]}\,p_{m,n}^\tau\,p_n^0\,,
\end{equation}
with $p_n^0$ being the eigenvalues of the initial density $\rho_0$, i.e. the initial occupation probabilities, and $p_{m,n}^\tau$  are the unitary transition probabilities, $p_{m,n}^\tau=\left|\bra{m(\alpha_\tau)} U_\tau \ket{n(\alpha_0)} \right|^2 $. It follows with $H_\mrm{H}(\alpha_\tau)$ being the Hamiltonian in the Heisenberg picture that \cite{kur00,tas00,tal07,cam11}
\begin{equation}
\label{q07}
\begin{split}
\la \e{-\beta W}\ra&=\la \e{-\beta H_\mrm{H}(\alpha_\tau)}\e{\beta H(\alpha_0)}\ra\\
&=\la \e{-\beta \Delta F}\ra\,,
\end{split}
\end{equation}
where $\la\dots\ra$ is here an average over the initial state $\rho_0$. Equation~\eqref{q07} led to the conclusion that quantum work is not an observable \cite{tal07}. The fundamental reason is that the Hamiltonian does not have to commute with itself at different times, $\com{H_\mrm{H}(\alpha_\tau)}{H(\alpha_0)}\neq 0$.

The classical and the  quantum Jarzynski equality are fundamentally different in their generality. Whereas the classical theorem holds true irrespective of the kind of dynamics, the quantum version \eqref{q07} relies on the unitarity of the dynamics. In addition, the stationary (equilibrium) state of general, open system dynamics are not described by a Gibbs state \cite{gelin_2009}. If we nevertheless follow the aforementioned quantum protocol and perform projective energy measurements, we have to account for the back action of the measurement of the system, that is 'the collapse of the wave function'. The fluctuation theorem has to be modified \cite{kafri_2012}, $\la \e{-\Delta E}\ra=\gamma$, where $\Delta E=E_m(\alpha_\tau)-E_n(\alpha_0)$ and $\gamma$ is the quantum efficacy quantifying the back action due to the projective measurement \cite{kafri_2012}. A distinction of work and heat is no longer feasible as by an energy measurement on the reduced system only changes in the internal energy can be determined. Moreover, the generalization of the entropy production for nonequilibrium stationary states \eqref{q04} is not obvious. In the next  section, we propose a new approach, with which these quantum issues can be resolved. 

\section{Fluctuation theorem in quantum phase space}

In the classical case the irreversible entropy production is defined along a path in phase space \eqref{q04}. The goal of the present paper is to generalize this concept to quantum phase space. However, various integral transformations have been proposed \cite{schleich_01} to represent the density operator $\rho_t$ in phase space. In the following it will prove convenient to work with the Wigner representation, as the time evolution equation for an open quantum system takes a particularly simple form \cite{cal83,hu92,garcia_2004}. The Wigner quasi probability distribution in phase space of a quantum state $\rho$ is given by \cite{wig32},
\begin{equation}
\label{q08}
\mc{W}_t(x,p)=\frac{1}{2\pi\hbar}\int \td y\,\e{-\frac{i}{\hbar}\,p y}\,\bra{x+\frac{y}{2}}\rho_t\ket{x-\frac{y}{2}}\,.
\end{equation}
The Wigner function contains the full classical information as its marginals are the probability distributions for the position $x$ and the momentum $p$, respectively. Note that generally $x$ and $p$ are vectors, whose dimension collects all degrees of freedom. However, $\mc{W}_t(x,p)$ is not a true probability distribution as $\mc{W}_t(x,p)$ can take negative values, which is a signature of quantum coherences \cite{schleich_01}.

The quantum Master equation can be written as
\begin{equation}
\label{q09}
\pd_t\,\mc{W}(\Gamma,t)=\mc{L}_{\alpha}\, \mc{W}(\Gamma,t)\,,
\end{equation}
where $\Gamma=(x,p)$ denotes again a point in phase space. It is worth emphasizing that a Liouvillian, $\mc{L}_\alpha$, does not generally exist for all quantum systems. In particular, for a thermally open harmonic oscillator it was shown in \cite{karrlein_1997} that existence and explicit form of $\mc{L}_\alpha$ are determined by the initial preparation of the environment. See also Ref.~\cite{breuer_07,fleming_2011} for further research on this topic.

The stationary solution of Eq.~\eqref{q09} is determined by
\begin{equation}
\label{q10}
\mc{L}_{\alpha}\,\mc{W}_\mrm{stat}(\Gamma,\alpha)=0\,.
\end{equation}
Generally the stationary Wigner function $\mc{W}_\mrm{stat}(\Gamma,\alpha)$ for an open quantum system in equilibrium is not given by the Wigner representation of the Gibbs state $\rho_\alpha=\e{-\beta H(\alpha}/Z(\alpha)$ \cite{for01,fleming_2011}, where $H(\alpha) $ is the Hamiltonian of the reduced system, only. Therefore, we conclude that the entropy production $\Sigma$, which fulfils a quantum equivalent of the classical fluctuation theorem, $\la\e{-\Sigma}\ra=1$, cannot simply be given by the classical expression of the thermodynamic work \eqref{q01}. We prove now that the true quantum entropy production $\Sigma$ rather reads 
\begin{equation}
\label{q11}
\Sigma[\Gamma_\tau;\,\alpha_\tau]=-\int_0^\tau \td t\,\dot\alpha_t\,\frac{\pd_\alpha\mc{W}_\mrm{stat}(\Gamma_t,\alpha_t)}{\mc{W}_\mrm{stat}(\Gamma_t,\alpha_t)}\,,
\end{equation}
which is the generalization of Eq.~\eqref{q13} to arbitrary coupling between the quantum system of interest and a thermal environment. Observe the equivalent forms of the quantum entropy production \eqref{q11} and the classical, nonequilibrium entropy production \eqref{q04}. Moreover, in Eq.~\eqref{q11} negative values of the Wigner function can be handled, without having to choose the correct Riemann surface of $\lo{\mc{W}}$.  

We emphasize that for the time being writing $\Sigma$ as a functional of a trajectory in phase space is a mathematical construct, which is convenient for the following proof. In the present context we understand the entropy produced along a quantum trajectory in analogy to Feynman path integrals \cite{zinn_2005}. Here a quantum trajectory is a mathematical tool defined as a generalization of the classical trajectory. Physical quantities are given by averages over an ensemble of such trajectories. In particular, we will see in the subsequent sections that the physically relevant observable $\mc{P}(\Sigma)$ can be completely determined from the dynamics described by Eq.~\eqref{q09}, without relying on 'single realizations' of $\Sigma$.

The proof is a straight forward generalization of the treatments proposed in \cite{jar97a} and \cite{speck_2007}. Consider the accumulated entropy $\sigma$ produced up to time $t$, $\sigma(t)=-\int_0^t \td s\,\dot\alpha_s\,\pd_\alpha\mc{W}_\mrm{stat}/\mc{W}_\mrm{stat}$, and thus $\sigma(\tau)=\Sigma$. Then the joint (quasi) probability distribution for the point in phase space and the accumulated entropy production, $P(\Gamma,\sigma,t)$, evolves according to,
\begin{equation}
\label{q12}
\pd_t\,P(\Gamma,\sigma,t)=\left[\mc{L}_\alpha-j_\mrm{stat}(\Gamma,\alpha_t)\,\pd_\sigma\right]\,P(\Gamma,\sigma,t)\,,
\end{equation}
where $j_\mrm{stat}(\Gamma,\alpha_t)$ is the (quasi) probability flux associated with the accumulated entropy production $\sigma$,
\begin{equation}
\label{q13}
j_\mrm{stat(\Gamma,\alpha_t)}=\dot\alpha_t\,\frac{\pd_\alpha\mc{W}_\mrm{stat}(\Gamma,\alpha_t)}{\mc{W}_\mrm{stat}(\Gamma,\alpha_t)}\,.
\end{equation}
Now we define the auxiliary density $\Psi(\Gamma,t)$ which is the exponentially weighted marginal of $P(\Gamma,\sigma,t)$. We have
\begin{equation}
\label{q14}
\Psi(\Gamma,t)=\int \td\sigma\,P(\Gamma,\sigma,t)\,\e{-\sigma}\,,
\end{equation}
for which the evolution equation \eqref{q12} becomes
\begin{equation}
\label{q15}
\pd_t\,\Psi(\Gamma,t)=\left[\mc{L}_\alpha-j_\mrm{stat}(\Gamma,\alpha_t)\right]\,\Psi(\Gamma,t)\,.
\end{equation}
It is easy to see that a solution of Eq.~\eqref{q15} is given by the stationary solution of the original master equation \eqref{q09} and we obtain
\begin{equation}
\label{q16}
\Psi(\Gamma,t)=\mc{W}_\mrm{stat}(\Gamma,\alpha_t)\,.
\end{equation}
Using the normalization of the stationary Wigner function \cite{schleich_01} we calculate with the latter solution for $\Psi(\Gamma,t)$,
\begin{equation}
\label{q17}
1=\int\td \Gamma\,\mc{W}_\mrm{stat}(\Gamma,\alpha_\tau)=\int\td \Gamma\,\Psi(\Gamma,\tau)=\la \e{-\Sigma}\ra\,,
\end{equation}
which concludes the proof. For any quantum system, open or closed, the entropy production fulfilling an integral fluctuation theorem is given by Eq.~\eqref{q11}. The only mathematical condition, which determines the range of validity, is the existence of a normalized stationary solution of Eq.~\eqref{q09}.

In the remainder of the paper we will discuss the physical significance of the latter mathematical finding. In particular, we will see how the entropy production defined in Eq.~\eqref{q11} is related to the thermodynamic work, and why the present approach is more useful for the theoretical understanding of the thermodynamic properties of open quantum systems than the approach based on two-time energy measurements.

\section{Illustration -- Harmonic oscillator}

As an illustration consider the time-dependent harmonic oscillator dragged through a thermal bath, whose bare Hamiltonian is given by
\begin{equation}
\label{q18}
H(x,p,t)=\frac{p^2}{2 m}+ V(x,t)=\frac{p^2}{2 m}+ \frac{1}{2} m \omega^2 (x- v t)^2\,,
\end{equation}
where $m$ is the mass, $\omega$ the angular frequency. For the sake of simplicity we choose the external control parameter $\alpha$ to be the position of the minimum of the oscillator, which is dragged with constant velocity $v$. The exact master equation for a time-dependent harmonic oscillator coupled to an environment consisting of an ensemble of harmonic oscillators is known \cite{hu92,zer95} and can be solved analytically \cite{for01,fleming_2011}. For simplicity let us consider a high temperature approximation of the exact evolution equation \cite{dil09}, for which the linear operator $\mc{L}_t$ in Eq.~\eqref{q09} reads in leading order of $\hbar$,
\begin{equation}
\label{q19}
\begin{split}
\mc{L}_t=-\frac{p}{m}\, \pd_x+V'(x,t)\,\pd_p\,+\pd_p\left(\gamma p+ D_{pp}\,\pd_p\right)+D_{xp}\,\pd_{xp}^2
\end{split}
\end{equation}
where $\gamma$ is the coupling coefficient to the environment,  $D_{pp}=m\gamma/\beta +m \beta\gamma\hbar^2(\omega^2-\gamma^2)/12$, and $D_{xp}=\beta\gamma\hbar^2/12$. As we mentioned earlier the master equation \eqref{q19} takes this particularly simple form only in the Wigner representation. Notice, for instance, that in the limit $\hbar\rightarrow 0$ Eq.~\eqref{q19} reduces to the classical Klein-Kramers equation \cite{risken_89}. The stationary solution can be written as
\begin{equation}
\label{q20}
\begin{split}
\mc{W}_\mrm{stat}(x,p)&=\frac{m \gamma\omega}{2\pi}\frac{1}{\sqrt{D_{pp}\left(D_{pp}+m\gamma\,D_{xp}\right)}}\\
&\times\e{-\frac{\gamma}{2}\left(\frac{p^2}{D_{pp}}+\frac{m^2\omega^2\,x^2}{D_{pp}+m\gamma\,D_{xp}}\right)}\,.
\end{split}
\end{equation}
Note that the latter $\mc{W}_\mrm{stat}(x,p)$ is not the Wigner representation of the Gibbs state, $\rho=\e{-\beta H(x,p,0)}/Z$, for the bare Hamiltonian $H(x,p,0)$ \eqref{q18}. The quantum entropy production \eqref{q11} reads,
\begin{equation}
\label{q21}
\Sigma[x_\tau,v]=\int_0^\tau \td t\, \frac{\gamma \omega^2 m^2}{D_{pp}+m\gamma\,D_{xp}}\,(x-vt)\, v\,.
\end{equation}
In the classical limit $\hbar\rightarrow 0$ the quantum entropy production \eqref{q21} becomes the classical irreversible work, $\Sigma\rightarrow\Sigma_\mrm{cl}=\int_0^\tau \td t\,\beta m\omega^2 \,(x-vt)\, v$, proposed in \cite{mazonka_99,zon03}.

We proceed with a numerical verification of the integral fluctuation theorem \eqref{q17} for the quantum entropy production \eqref{q11}. To this end, we numerically integrated Eq.~\eqref{q12} with the linear operator \eqref{q19} and the stationary solution \eqref{q20}. The probability distribution of the quantum entropy production is given by the marginal of $P(\Gamma,\sigma,t)$ at the end of the driving interval of length $t=\tau$, $\mc{P}(\Sigma)=\int\td\Gamma\,P(\Gamma,\Sigma,\tau)$. Since the stationary solution \eqref{q20} is Gaussian and the driving linear \eqref{q18} we expect in analogy to the classical case \cite{mazonka_99,zon03} $\mc{P}(\Sigma)$ to be Gaussian, as well.
\begin{figure}
\includegraphics[width=0.45\textwidth]{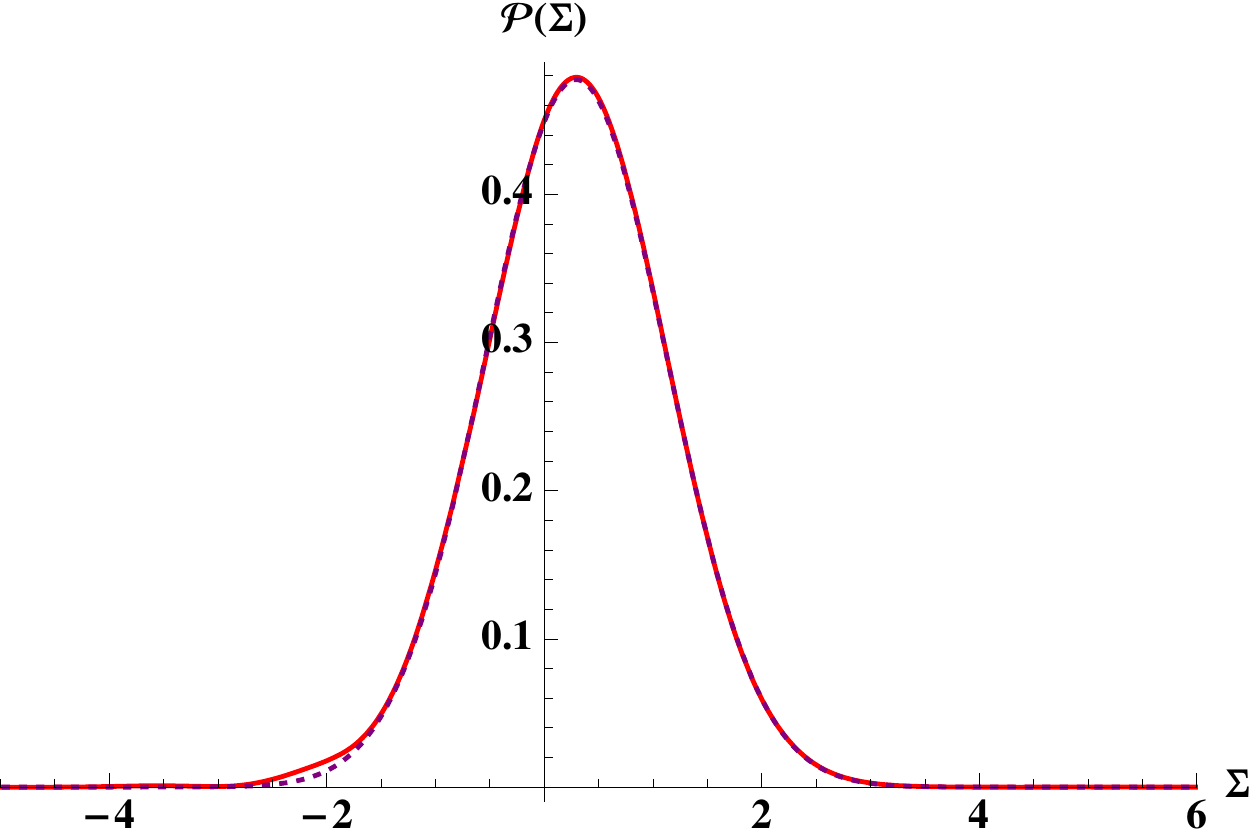}
\caption{(color online)\label{fig1} Probability distribution $\mc{P}(\Sigma)$ (red, solid line) together with a Gaussian fit (purple, dotted line) for $m=1$, $\hbar\omega=0.1$, $\beta=1$, $\tau=1$, $\gamma=1$ and $v=0.8$.}
\end{figure}
In Fig.~\ref{fig1} we plot $\mc{P}(\Sigma)$ for one set of parameters together with a Gaussian fit. In Fig.~\ref{fig2} we plot the numerical check of resulting fluctuation theorem \eqref{q17} as a function of the driving time $\tau$. We observe that Eq.~\eqref{q17} is verified to very high numerical precision  \footnote[1]{The aberrations for larger values of $\tau$ are a numerical artefact, stemming from the limited precision of the numerical integration with Wolfram Mathematica 9.0.1.0.}.
\begin{figure}
\includegraphics[width=0.45\textwidth]{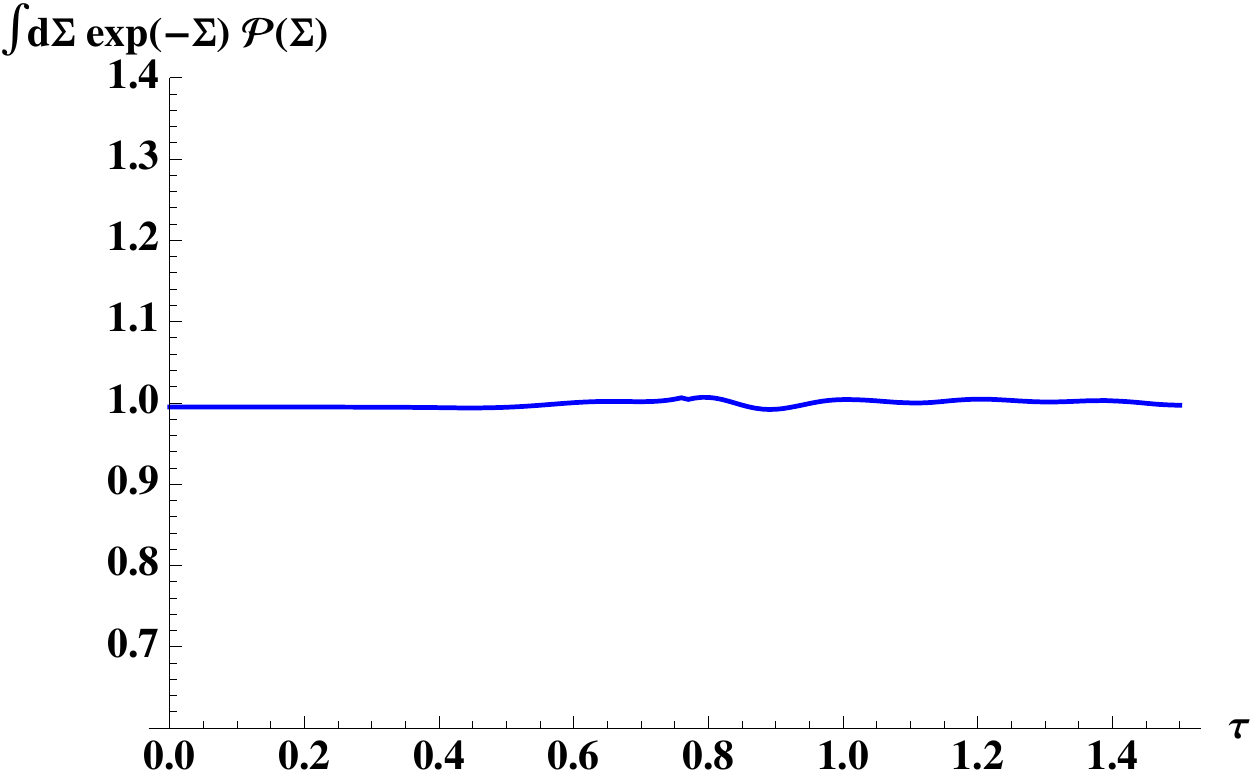}
\caption{(color online)\label{fig2} Numerical check of the fluctuation theorem \eqref{q17} with parameters of Fig.~\ref{fig1} as a function of $\tau$.}
\end{figure}

\section{Significance of the new approach}

In the preceding sections we derived a fluctuation theorem \eqref{q17} for the entropy production in Wigner phase space \eqref{q11} and illustrated the general findings with the help of the harmonic oscillator dragged trough a thermal bath \eqref{q18}. In the following, we will explain how, at least in principle, the quantum entropy production \eqref{q11} could  be determined in an experiment. Furthermore, we will demonstrate that the new formulation \eqref{q17} is capable of explaining physical situations, which are beyond the scope of the two time measurement approach. 

\subsection{Experimental verifications}

From the definition of the entropy production \eqref{q11} one might be tempted to consider the entropy produced along a path in phase space. Whereas this concept makes sense for classical systems, for quantum systems this interpretation has to be treated with utmost care. Single realizations of the entropy production \eqref{q11} should be rather interpreted like trajectories in the path integral formalism \cite{zinn_2005}. These are mathematical constructs developed for describing a quantum system with classical means. The actual physical quantities are given by averages over ensembles of realizations, or trajectories, which is in the present context is e.g. the probability distribution $\mc{P}(\Sigma)=\la \de{\Sigma-\Sigma[\Gamma_\tau;\,\alpha_\tau]}\ra$.

Having this in mind one could argue that the two time measurement approach is the more 'natural' treatment of quantum thermodynamic processes. However, different proposals for realistic experimental verifications of the quantum Jarzynski equality \eqref{q07} were so far mostly restricted to isolated quantum systems consisting of harmonic oscillators \cite{hub08,dorner_2013}. In principle, the measurements procedures proposed in \cite{hub08,dorner_2013} could be extended to more general systems. However, in order to measure $\mc{P}_\mrm{qm}(W)$ \eqref{q06} the energy eigenvalues \eqref{q05} of the system under consideration have to be determined first.

Experimental verifications of the fluctuation theorem for the quantum entropy production in phase space are not necessarily easier to achieve. In principle, one would have to measure, the entropic flux \eqref{q13}. To this end, one has to determine the stationary Wigner function for different values of $\alpha$. This can be done by preparing the quantum system in one configuration, that is one specific value of $\alpha$, let the quantum system relax into its stationary state, and then measure the Wigner function with the help of quantum tomography \cite{nielsen_00} or directly \cite{bertet_2002}. This procedure has to be repeated for all values of $\alpha$. As a second quantity the linear operator $\mc{L}_\alpha$ has to be determined. Finally, $\mc{P}(\Sigma)$ is numerically determined as illustrated for the harmonic oscillator in the preceding section. This last step is equivalent to constructing $\mc{P}_\mrm{qm}(W)$ \eqref{q06} from the experimentally determined transition probabilities $p_{m,n}^\tau$ \cite{hub08}.

The main advantage of the present approach is its universality as open and isolated systems, in or far from thermal equilibrium, are described by the same means.

\subsection{Nonequilibrium stationary states}

As an illustration we discussed the harmonic oscillator dragged trough a thermal bath, for which the stationary solution is an equilibrium state. However, the present approach can also be applied to situations where the stationary state is far from equilibrium, as, for instance, even in the long time limit non-vanishing fluxes persist. These physical situations are generally described by evolution equations of the form \eqref{q19} and including non-conservative forces. The fluctuation theorem \eqref{q17} for the entropy production in phase space \eqref{q11} generalizes the classical Hatano-Sasa relation \cite{hat01} in a natural way to open quantum systems. 

\subsection{Microcanonical ensembles}

Within the two time energy measurement approach the fluctuation theorem, and hence an expression for the entropy production was proposed for microcanonical ensembles \cite{talkner_2008,talkner_2013}. It is worth emphasizing that the present approach can also capture these physical situations. The above introduced entropy production \eqref{q11} is given entirely in terms of the stationary solution, no matter what ensemble is describe by $\mc{W}_\mrm{stat}$. Hence, the fluctuation theorem \eqref{q17} is universally valid, and determines the physical nonequilibrium entropy prodcution for microcanonical, canonical, grandcanonical, etc. ensembles. 

\subsection{Semiclassical approximations}

Generally, the derivation of quantum evolution equations like Eq.~\eqref{q19} is mathematically involved. Therefore, one commonly invokes high temperature or semiclassical approximations \cite{breuer_07}. A particularly interesting case is the limit of high damping, $\hbar\beta\gamma\gg 1$. Starting from the exact master equation \cite{hu92} a semiclassical Smoluchowski equation can be derived \cite{dil09}, which takes the form
\begin{equation}
\label{q23}
\pd_t p(x,t)=\frac{1}{\gamma\,m}\,\pd_x\left[V'(x,t)+\frac{1}{\beta}\,D_e(x,t)\right]\,p(x,t)\,,
\end{equation}
where $D_e(x,t)=1/[1-\lambda\beta V''(x,t)]$, and $\lambda(\hbar)$ is the quantum parameter, which depends non-trivially on $\hbar$. See also \cite{ank01,ank08,mac04,ank05,mai10,cof07} for alternative derivations. The stationary solution reads,
\begin{equation}
\label{q24}
p_\mrm{stat}(x,\alpha)=\frac{1}{Z(\alpha)}\,\frac{\e{-\beta V(x,\alpha)+\lambda \beta^2 {V'(x,\alpha)}^2/2}}{1-\lambda\beta V''(x,\alpha)}\,,
\end{equation}
where $V(x,\alpha)$ is the potential. In \cite{deffner_2011} we showed that then a fluctuation theorem can be derived, $\la \e{-\Sigma_\mrm{QSE}}\ra=1$, where $\Sigma_\mrm{QSE}[x_\tau;\,\alpha_\tau]=\int_0^\tau \td t\,\dot\alpha_t\,\pd_\alpha \lo{p_\mrm{stat}(x_t,\alpha_t)}$. Moreover, we showed in \cite{deffner_2011} that $\Sigma_\mrm{QSE}$ is given by the classical entropy production plus correction terms accounting for the quantum fluctuations. This result was  also discovered in \cite{subasi_2012} for quantum Brownian motion. 

These results can be naturally obtained from the above introduced expression for the entropy production \eqref{q11} and its according fluctuation theorem \eqref{q17} by taking the appropriate limit, $\hbar\beta\gamma\gg1$. Thus, we conclude that the present treatment can serve as a starting point for various semiclassical approximations.

\section{Concluding remarks}

In the present paper a new approach to describe the thermodynamic properties of open quantum systems has been proposed. Motivated by classical formulations of the entropy production and its corresponding fluctuation theorem the quantum entropy production has been defined as a functional of a trajectory in Wigner phase space. In analogy to the path integral formalism this trajectory based approach is to be understood as a mathematical tool to define and understand physical quantities, which are given by averages over ensembles of these trajectories, or 'single realizations' of physical processes. In the present case we have been interested in the probability distribution of the entropy production $\mc{P}(\Sigma)$, whose mean $\la \Sigma\ra$ is the physically significant quantity. The new approach has been elucidated by discussing its experimental accessibility and its universality. Finally, the phase space treatment has been opposed to the two time energy measurement approach, where the present phase space approach has been shown to be more general. 

As a final remark, we note that in Refs.~\cite{horowitz_2012,campisi_2013} others approaches were proposed, which do not rely on two time measurements.

\acknowledgments{It is a pleasure to thank Zhiyue Lu for stimulating discussions. We acknowledge financial support by a fellowship within the postdoc-program of the German Academic Exchange Service (DAAD, contract No D/11/40955) and from the National Science Foundation (USA) under grant DMR-1206971}.

\bibliographystyle{eplbib}

\end{document}